\documentclass[reprint, 10pt, a4paper, aip, floatfix, longbibliography]{revtex4-2}
\usepackage[utf8]{inputenc}
\usepackage{amsmath}
\usepackage{amsfonts}
\usepackage{amssymb}
\usepackage{mathrsfs}
\usepackage{subfigure}
\usepackage[hidelinks]{hyperref}
\usepackage{outlines}
\usepackage{tikz}
\usepackage{booktabs}

\newcommand{\bv}{\mathbf{v}}
\newcommand{\bu}{\mathbf{u}}
\newcommand{\bE}{\mathbf{E}}
\newcommand{\bB}{\mathbf{B}}
\newcommand{\bF}{\mathbf{F}}

\newcommand{\D}{\mathrm{d}}

\newcommand{\rot}{\omega_\text{rot}}

\begin{document}

\title{Coriolis Forces Modify Magnetostatic Ponderomotive Potentials}
\date{\today}
\author{E.~J. Kolmes}
\email[Electronic mail: ]{ekolmes@princeton.edu}
\affiliation{Department of Astrophysical Sciences, Princeton University, Princeton, New Jersey 08544, USA}
\author{N.~J. Fisch}
\affiliation{Department of Astrophysical Sciences, Princeton University, Princeton, New Jersey 08544, USA}

\begin{abstract}
It is possible to produce a ponderomotive effect in a plasma system without time-varying fields, if the plasma flows over spatial oscillations in the field. 
This can be achieved by superimposing a spatially oscillatory perturbation on a guide field, then setting up an electric field perpendicular to the guide field to drive flow over the perturbation. 
However, subtle distinctions in the structure of the resulting electric field can entirely change the behavior of the resulting ponderomotive force. 
Previous work has shown that, in slab models, these distinctions can be explained in terms of the polarization of the effective wave that appears in the co-moving frame. 
Here we consider what happens to this picture in a cylindrical system, where the transformation to the co-moving (rotating) frame is not inertial. 
It turns out that the non-inertial nature of this frame transformation can lead to counterintuitive behavior, partly due to the appearance of parallel (magnetic-field-aligned) electric fields in the rotating frame even in cases where none existed in the laboratory frame. 
Apart from the academic interest of this study, the practical impact lies in being better able to anticipate the antenna configuration on the plasma periphery of a cylindrical plasma that will lead to optimal ponderomotive barrier formation in the interior plasma.
\end{abstract}

\maketitle

\section{Introduction} \label{sec:introduction}

In the ponderomotive effect, a charged particle subject to an inhomogeneous wave field is repelled from or attracted to the region where the field is strongest. 
In the context of magnetic confinement for plasmas, there is longstanding interest in the possible use of ponderomotive forces to help trap plasmas.
For plasma traps, the usual approach is to use an RF antenna to inject an electromagnetic wave into a particular region so as to better trap ions. \cite{Watari1978, Fader1981, Phaedrus1985, Miller2023}
This could be very helpful for improving confinement times for fusion devices. 
Much the same physics can be applied for applications involving plasma acceleration.\cite{Motz1967}
For some applications -- for example, aneutronic fusion\cite{Magee2019, Putvinski2019, Kolmes2022PowerBalance, Ochs2022, Magee2023} -- it may be of particular interest that these ponderomotive effects can selectively confine or deconfine different particle species.\cite{Rubin2023, Rubin2023ii, Ochs2023Isopotentials, Rubin2024} 
Ponderomotive forces can also be used to stabilize otherwise unstable modes in magnetic confinement devices.\cite{Ferron1983, Yasaka1986, Browning1988, Browning1989} 

Although it is usually understood in the context of RF waves, the same effect also appears when a plasma flows through \textit{spatially} oscillating fields. 
Such a configuration could have major advantages. 
Many plasma configurations have some tendency to flow (especially to rotate).\cite{Lehnert1971, Suckewer1979, Bekhtenev1980, Ellis2001, Helander2008, Endrizzi2023}
If the plasma is already flowing, it may be simpler and more efficient to impose static perturbations to the fields than to inject RF waves. 

For example, consider a configuration with a constant axial magnetic field and a radial electric field. 
These perpendicular fields produce $\bE \times \bB$ rotation. 
Then imagine adding a multipole perturbation to the magnetic field. 
The resulting periodic azimuthal structure is sufficient to produce a ponderomotive effect without the use of any time-varying fields. 
This is possible because as a particle rotates, it samples different field regions so that it experiences a local field that oscillates as a function of azimuthal position. 

This configuration was suggested by \citet{Rubin2023, Rubin2023ii}, who used Hamiltonian methods to show that it leads to a ponderomotive barrier. 
Their analysis assumed that the electric field was unchanged by the introduction of the magnetostatic perturbation, so that the \textit{unperturbed} magnetic flux surfaces are the isopotential surfaces. 

\begin{figure}
\includegraphics[width=\linewidth]{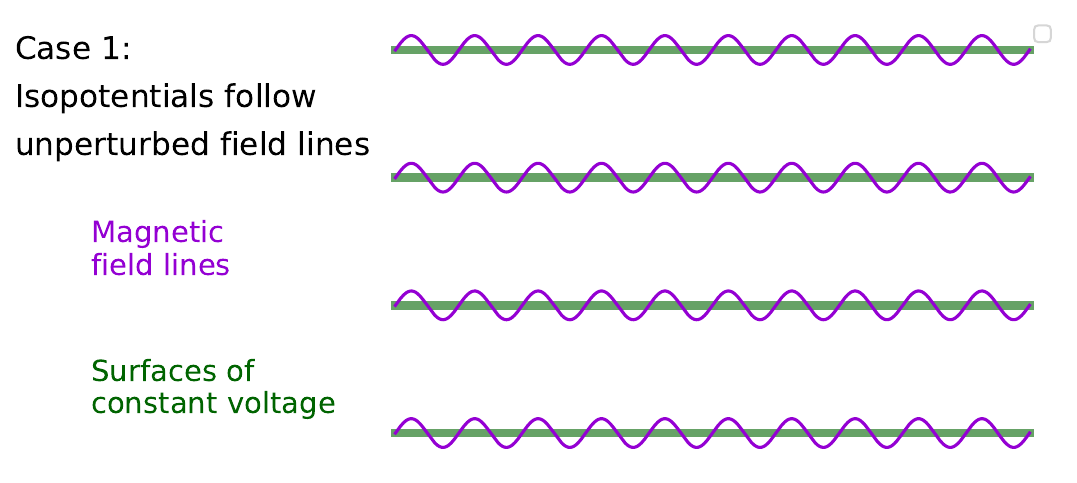} \\ ~\\
\includegraphics[width=\linewidth]{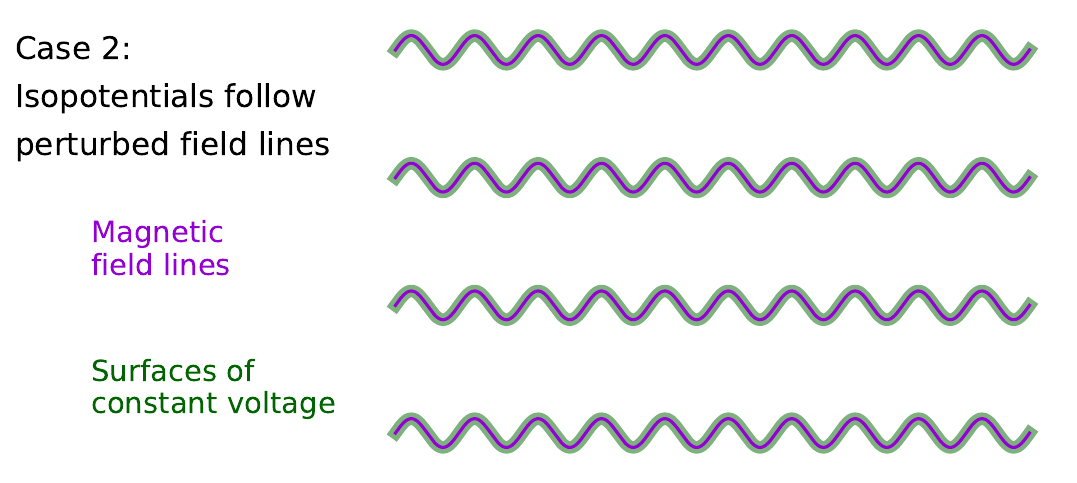}
\caption{This cartoon illustrates the difference between two possible choices of electric field when the magnetic field is a straight guide field with a small perturbation. In the first case, the surfaces of constant voltage (green) follow the unperturbed magnetic field lines (violet). In the second, they follow the perturbed field lines (so that the green and violet sit on top of one another). }
\label{fig:fieldCartoon}

\end{figure}

\citet{Ochs2023Isopotentials} studied a magnetostatic ponderomotive configuration in a related slab system. 
They showed that the choice between perturbed and unperturbed flux surfaces as isopotentials entirely changes the behavior of the ponderomotive potential. 
The distinction between perturbed and unperturbed flux surface isopotentials is illustrated in Figure~\ref{fig:fieldCartoon}. 
Their approach relied on a transformation to the frame moving at the $\bE\times\bB$ velocity. 
In the moving frame, the presence or absence of small electric fields parallel to the full (perturbed) magnetic field changes the polarization of the effective RF wave experienced by the particle as it drifts across the field oscillations. 

The analysis -- and much of the intuition -- behind the work by \citet{Ochs2023Isopotentials} rests on the mapping between the static perturbed fields and the effective RF wave experienced by the particle in the co-moving frame (accomplished formally using a Lorentz transformation). 
This leads to the question: what happens when we try to apply the same ideas to a cylindrical system like the one originally considered by \citet{Rubin2023, Rubin2023ii}? 
It is still possible to consider a transformation to a co-moving frame, but the resulting rotating frame is not inertial and behaves differently in a number of ways. 
These include modifications of the ponderomotive force as well as non-ponderomotive effects that can compete with the ponderomotive force. 

This paper is organized as follows. 
Section~\ref{sec:polarization} reviews the relationship between the polarization of a wave and the ponderomotive potential it produces. 
Section~\ref{sec:frame} describes rotating-frame transformations and their implications, including a modification of the resonance conditions and the appearance of parallel fields in the rotating frame even when there are none in the laboratory frame. 
Section~\ref{sec:multipoleExample} shows how these effects play out in the case of the multipole fields considered by \citet{Rubin2023, Rubin2023ii}, but with electric fields modified using the assumption that the perturbed magnetic flux surfaces are surfaces of constant potential. 
Section~\ref{sec:discussion} is a discussion of these results. 

\section{The Role of Polarization} \label{sec:polarization}

\begin{figure*}
	\includegraphics[width=.48\linewidth]{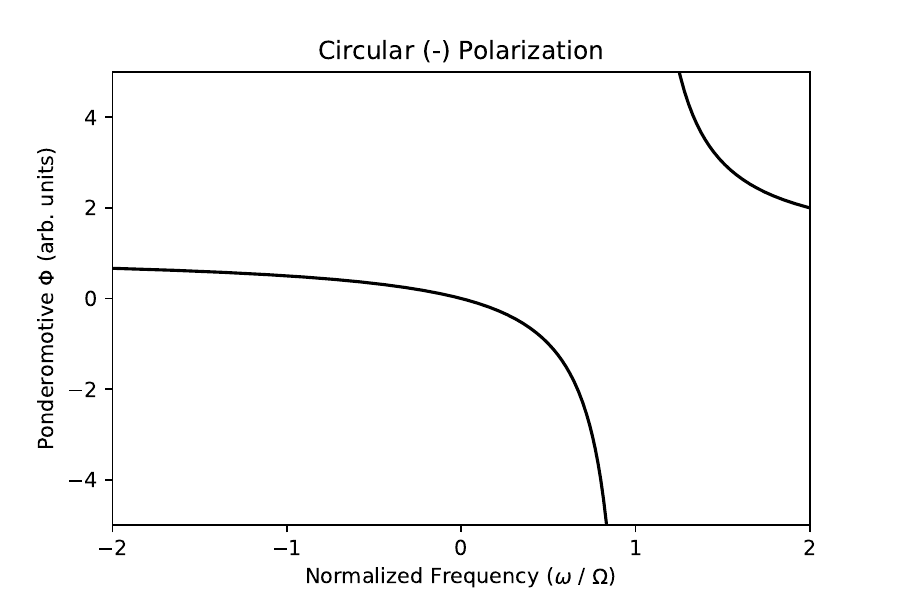} 
	\includegraphics[width=.48\linewidth]{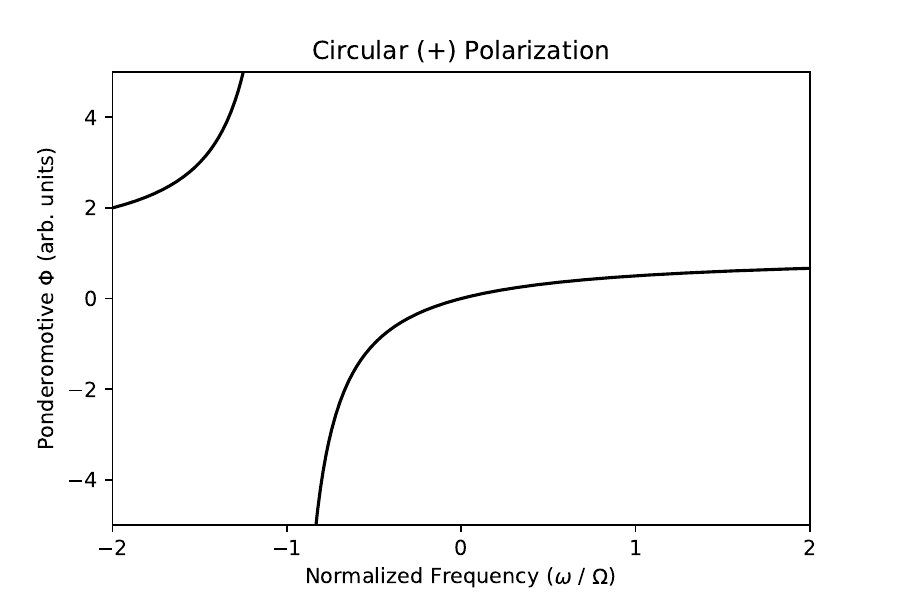} \\
	\includegraphics[width=.48\linewidth]{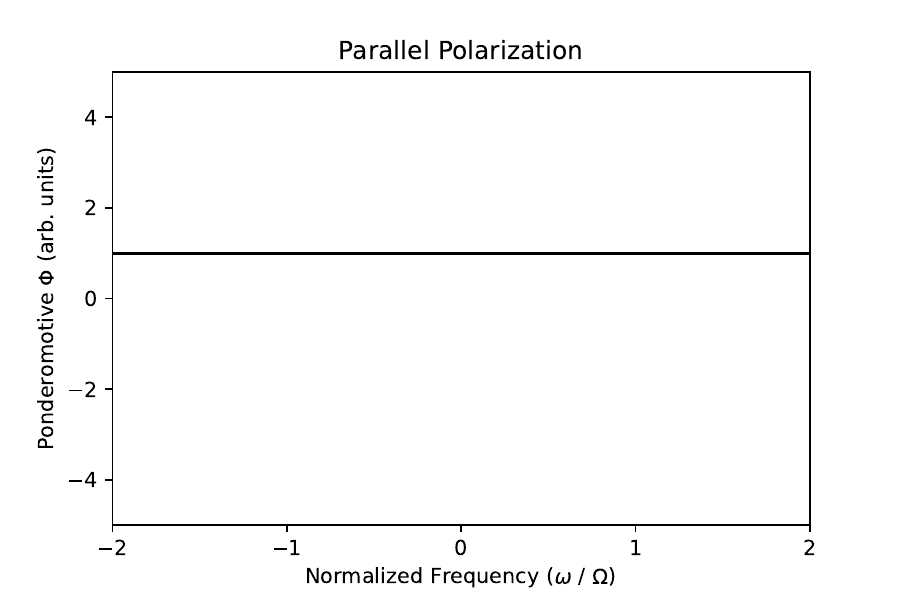}
	\includegraphics[width=.48\linewidth]{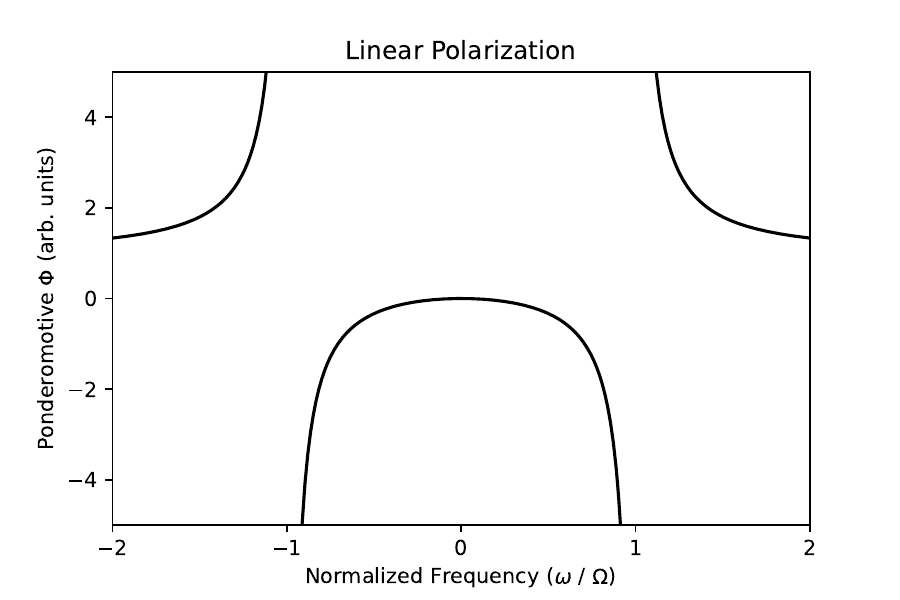}
	\caption{The qualitative structures of ponderomotive potentials associated with four different polarizations. The linear polarization is a combination of equal parts of the two circular polarizations. In all four cases, the amplitude of the electric field is assumed to be proportional to $\omega$. Note that, formally, either $\omega$ or $\Omega$ can be negative; $\Omega$ is the signed gyrofrequency and therefore depends on the sign of the particle's charge, and the polarizations are defined in a way that allows the wave frequency to be negative (even though it might sometimes be simpler to map a negative-frequency wave with one circular polarization onto a positive-frequency wave with the other). } 
	\label{fig:polarizationStructures}
\end{figure*}

One way of deriving the ponderomotive force is to split the particle dynamics into fast and slow motion. 
Suppose a magnetized particle interacts with some oscillating (RF) electromagnetic field, and suppose the particle moves between regions of higher and lower oscillating fields on timescales much slower than those of the oscillations. 
Then the slow dynamics describes the motion of the particle gyro-center through the wave fields and the fast dynamics describes the oscillations produced by interactions with the RF fields. 
The resulting motion approximately conserves the magnetic moment $\mu \doteq m v_{f,\perp}^2 / 2 B$, where $m$ is the particle mass, $v_{f,\perp}$ is the perpendicular velocity not including fast RF-driven oscillations, and $B$ is the background magnetic field strength (which we will take to be large compared with any oscillating component). 
The motion also approximately conserves the ``quasi-energy"\cite{Dodin2005Integrals} 
\begin{gather}
\mathcal{E} \doteq \frac{m}{2} \, \big\langle v_{||} \big \rangle^2 + \mu B + \Phi, \label{eqn:quasiEnergy}
\end{gather}
where $\langle v_{||} \rangle$ is the parallel velocity averaged over the RF-driven oscillations and $\Phi$ is the ponderomotive potential. 

The behavior of this ponderomotive potential is strongly dependent on the polarization of the RF wave. 
The applied electric field can be decomposed into a component parallel to the magnetic field, denoted by $E_{||}$, and circularly polarized components perpendicular to the field, given by\cite{Dodin2005Integrals} 
\begin{align}
E_+ &\doteq \frac{1}{\sqrt{2}} \big( E_x - i E_y \big) \label{eqn:EPlus} \\
E_- &\doteq \frac{1}{\sqrt{2}} \big( E_x + i E_y \big) \label{eqn:EMinus}, 
\end{align}
where $(x,y,z)$ are the usual Cartesian coordinates and the background magnetic field is in the $z$ direction. 
In a cold plasma, the resulting ponderomotive potential can be written as\cite{Motz1967, Ferron1983, Browning1988, Browning1989, Dodin2005Integrals} 
\begin{gather}
\Phi = \frac{q^2}{4 m \omega^2} \bigg( \frac{|E_+|^2}{1 + \Omega / \omega} + \frac{|E_-|^2}{1 - \Omega / \omega} + |E_{||}|^2 \bigg) , \label{eqn:ponderomotivePhi}
\end{gather} 
where $q$ is the charge of the particle, $\omega$ is the frequency of the wave, and $\Omega \doteq q B / m$ is the gyrofrequency. 
Eq.~(\ref{eqn:ponderomotivePhi}) breaks down near the singularities where $\Phi$ would otherwise diverge.\cite{Dodin2004} 
Formally, Eq.~(\ref{eqn:ponderomotivePhi}) can be used in a system with plasma flows by transforming to the co-moving (possibly non-inertial) frame. 
Fields with spatial oscillations in the laboratory frame can become time-varying in the co-moving frame. 

As is shown in Eq.~(\ref{eqn:ponderomotivePhi}), differently polarized components of $\bE$ contribute very differently to the ponderomotive potential. 
This is illustrated in Figure~\ref{fig:polarizationStructures}. 
If a small change to the fields modifies this polarization, it can change the magnitude and even the sign of $\Phi$. 
\citet{Ochs2023Isopotentials} found that, in a slab system related to the wiggler considered by \citet{Rubin2023}, shifting the isopotential surfaces from the unperturbed magnetic flux surfaces to the perturbed flux surfaces changed the polarization from dominantly parallel to transversely (linearly) polarized (that is, from $E_{||}$ to equal parts $E_+$ and $E_-$). 
This polarization picture can be very convenient, because determining the polarization of $\bE$ in the co-moving frame is typically much simpler than calculating the dynamics from scratch. 
It provides a framework with which to understand the effects of whole classes of fields, without the need to recalculate the particle dynamics for each case. 

\section{Rotating-Frame Transformations} \label{sec:frame}

Formally, one way to calculate the behavior of the effective ponderomotive potential in a flowing plasma is to transform to the co-moving frame that eliminates the flow, then to calculate the decomposition of the electric field in that frame into $E_+$, $E_-$, and $E_{||}$. 
The resulting ponderomotive potential can be inferred from this decomposition in the same way described in the previous section. 
Throughout this paper, we will take drift velocities to be non-relativistic. 

Consider a plasma with laboratory-frame electric and magnetic fields $\bE$ and $\bB$. 
In the frame moving at (non-relativistic) velocity $\bu$, the dynamics are identical to what we would see at rest with effective fields\cite{Thyagaraja2009} 
\begin{align}
	\bE_* &= \bE + \bu \times \bB - \frac{m}{q} \frac{\partial \bu}{\partial t} + \frac{m}{2 q} \nabla u^2 \\
	\bB_* &= \bB + \frac{m}{q} \nabla \times \bu . 
\end{align}
Consider the solid-body rotating frame in which $\bu = r \omega_\text{rot} \hat \theta$. 
Take $(r, \theta, z)$ to be the standard cylindrical coordinates, with corresponding unit vectors $\hat r$, $\hat \theta$, and $\hat z$. 
The fields in this frame are 
\begin{align}
	\bE_* &= \bE + r \omega_\text{rot} \hat \theta \times \bB + \frac{m r \omega_\text{rot}^2}{q} \hat r \\
	\bB_* &= \bB + \frac{2 m \omega_\text{rot}}{q} \hat z . \label{eqn:BStar}
\end{align}
Now suppose $\bB = B_0 \hat z + \bB_1$ and $\bE = (r/R) E_0 \hat r + \bE_1$, with $B_0$ and $E_0$ taken to be constant for the time being. 

Let $\omega_\text{rot}$ correspond to the frequency of the Brillouin slow mode in absence of $\bE_1$ and $\bB_1$: 
\begin{align}
	\omega_\text{rot} = - \frac{q B_0}{2 m} \bigg( 1 - \sqrt{ 1 - 4 \frac{mE_0}{qRB_0^2} } \, \bigg) . \label{eqn:slowBrillouin}
\end{align}
Define the unperturbed $\bE \times \bB$ frequency by 
\begin{gather}
	\omega_E \doteq - \frac{E_0}{R B_0} 
\end{gather}
and let $\Omega_0 \doteq q B_0 / m$. 
Note that 
\begin{gather}
	\rot = \omega_E \bigg[ 1 - \frac{\omega_E}{\Omega_0} + \mathcal{O} \bigg( \frac{\omega_E^2}{\Omega_0^2} \bigg) \bigg] . 
\end{gather}
$\rot$ tends toward $\omega_E$ when the unperturbed $\bE \times \bB$ frequency is small compared with $\Omega_0$. 
In the frame moving at $\bu = r \omega_\text{rot} \hat \theta$, the effective fields are 
\begin{align}
	\bE_* 
	&= \bE_1 + r \rot \hat \theta \times \bB_1 
\end{align}
and 
\begin{align}
	\bB_* &= \bB_0 + \bB_1 + \frac{2 m \rot}{q} \, \hat z . 
\end{align}
The slab limit can be recovered by taking $\rot \rightarrow 0$ and $r \rightarrow \infty$ while holding $r \rot$ constant. 

One key difference between the rotating-frame and translating-frame transformations is that in the former, $\bB_*$ is modified inertially by the Coriolis force even for non-relativistic flow velocities, whereas in the latter $\bB_*$ changes only in the relativistic limit. 
This modifies the resonance conditions that appear for the different polarizations from Eq.~(\ref{eqn:ponderomotivePhi}). 
Consider a frame rotating at angular frequency $\rot$, and suppose there are static, azimuthal field perturbations with azimuthal mode number $n$ -- that is, perturbing fields that vary azimuthally like $\sin(n \theta)$ and $\cos(n \theta)$. 
The resulting wave in the rotating frame has frequency $n \rot$, and the ponderomotive potential is 
\begin{align}
\Phi_* &= \frac{q^2}{4 m n^2 \rot^2} \bigg( \frac{n \rot |E_{*+}|^2}{(n+2)\rot + \Omega_0} \nonumber \\
&\hspace{60pt} + \frac{n \rot |E_{*-}|^2}{(n-2) \rot - \Omega_0} + |E_{*||}|^2 \bigg), \label{eqn:ponderomotivePhiStar}
\end{align}
where the starred field components are evaluated in the rotating frame. 
The appearance of $n+2$ and $n-2$ in the denominators (rather than $n$) comes from the inertial modification of the effective magnetic field. 
This shifts the resonances. 
It also renders the resonance for the $n=2$ negative circular polarization inaccessible. 
This is because the $n=2$ resonance condition for $E_{*-}$ requires $2 \rot$ to match the gyrofrequency in the rotating frame, but the rotating-frame gyrofrequency is $\Omega_0 + 2 \rot$. As $\rot$ increases, the gyrofrequency ``outruns" the frequency of the wave. 

Another key qualitative difference between the slab and cylindrical cases has to do with the parallel component of $\bE$. 
In a uniformly flowing slab, the appropriate frame transformation is the usual translation. 
$\bE \cdot \bB$ is a Lorentz invariant. 
As such, if $\bE$ is perpendicular to $\bB$ in the laboratory frame, the two fields will remain perpendicular in the co-moving frame. 
This means that any time the magnetic field lines are isopotential surfaces in the laboratory frame, the parallel polarization will not enter into the ponderomotive potential. 

However, the rotating-frame transformation is \textit{not} an inertial transformation, and does not in general preserve Lorentz invariants. 
As we will see, this means that the parallel polarization can appear in the ponderomotive potential for a rotating plasma even if $\bE$ and $\bB$ are entirely perpendicular in the laboratory frame.  
\begin{align}
\bE_* \cdot \bB_* 
&= \bE \cdot \bB \bigg(1 + \frac{2 \rot}{\Omega_0} \bigg) - \frac{2 \rot}{\Omega_0} \bE_1 \cdot \bB_1 \nonumber \\
&\hspace{10 pt}+ r B_0 B_{1r} ( \omega_E - \rot ) \bigg(1 + \frac{2 \rot}{\Omega_0} \bigg) . 
\end{align}
If $\bE \cdot \bB = 0$ -- that is, if field lines are isopotentials in the laboratory frame -- then this is 
\begin{align}
\bE_* \cdot \bB_* &= r B_0 B_{1r} (\omega_E - \rot) \bigg( 1 + \frac{2 \rot}{\Omega_0} \bigg) \nonumber \\
&\hspace{50 pt}- \frac{2 \rot}{\Omega_0} \bE_1 \cdot \bB_1. \label{eqn:EStarDotBStar}
\end{align}
In the limit where $\omega_E \ll \Omega_0$, this can be written to leading order in $\omega_E / \Omega_0$ as 
\begin{align}
\bE_* \cdot \bB_* &= - \bigg( \frac{r E_0 \hat r}{R} \cdot \bB_1 + 2 \bE_1 \cdot \bB_1 \bigg) \frac{\omega_E}{\Omega_0} \,  . \label{eqn:lowFlowEStarDotBStar}
\end{align}
The key result of this calculation is the surprise appearance of a parallel component of $\bE_*$ even in cases where the laboratory-frame $\bE$ is purely perpendicular to $\bB$. 
(In principle, the converse can also occur, with the parallel field vanishing in the rotating frame but nonvanishing in the laboratory frame, though this would require very carefully chosen fields.) 
$\omega_E / \Omega_0$ is the dimensionless parameter that controls the importance of inertial effects like the centrifugal and Coriolis forces. 
When this parameter is small, $\omega_E$ and $\rot$ become indistinguishable and we recover the slab limit. 
Physically, the discrepancy between $\omega_E$ and $\rot$ comes from the centrifugal $\bF \times \bB$ drift, which cannot compete with the $\bE\times\bB$ drift when $\omega_E \ll \Omega_0$. 


\section{Multipole Fields} \label{sec:multipoleExample}

Some of these effects may be easier to understand with the help of a worked example. 
To that end, consider the magnetic field configuration from \citet{Rubin2023, Rubin2023ii}, which included an axial guide field and a multipole perturbation. 
Consider a generalization of that field in which the guide field is also allowed to vary axially: 
\begin{align}
&\bB = B_0(z) \hat z + B_1(z) \bigg( \frac{r}{R} \bigg)^{n-1} \big[ \cos (n \theta) \hat r - \sin (n \theta) \hat \theta \, \big] \nonumber \\
&\hspace{120 pt}- \frac{r}{2} B_0'(z) \hat r ,  \label{eqn:multipoleB}
\end{align}
where $B_1(z) \ll B_0(z)$, $R$ is a constant, $n$ is a positive integer, $(r, \theta, z)$ are the usual cylindrical coordinates, and $\hat r$, $\hat \theta$, $\hat z$ are the corresponding unit vectors. 
$B_0(z)$ is assumed to be nonvanishing everywhere. 

If we assume that the unperturbed flux surfaces are surfaces of constant electrical potential, then the electrostatic potential $\varphi_\text{un}$ is a function of $r$ and $B_0(z)$. 
Let 
\begin{gather}
\varphi_\text{un}(r,\theta,z) = - \frac{E_0 r^2}{2 R} \frac{B_0(z)}{B_0(z_0)} \, , 
\end{gather}
since this results in solid-body rotation at leading order in $B_1 / B_0$. 
If we instead assume that the perturbed flux surfaces are isopotentials, then the electrostatic potential $\varphi_\text{iso}$ that corresponds to $\varphi_\text{un}$ at $z = z_0$ is defined by the requirements that 
\begin{gather}
\varphi_\text{iso}(r,\theta,z_0) = - \frac{E_0 r^2}{2 R} 
\end{gather}
and 
\begin{gather}
B_r \frac{\partial \varphi_\text{iso}}{\partial r} + \frac{B_\theta}{r} \frac{\partial \varphi_\text{iso}}{\partial \theta} + B_z \frac{\partial \varphi_\text{iso}}{\partial z} = 0. \label{eqn:isopotentialCurve}
\end{gather}
Eq.~(\ref{eqn:isopotentialCurve}) is precisely the requirement that $\varphi_\text{iso}$ be constant along field lines. 
$\varphi_\text{iso}$ can be calculated exactly in some cases and approximately in others. 

\subsection{Quadrupole Fields}

First consider the case where $n = 2$. 
Then Eq.~(\ref{eqn:multipoleB}) can be expressed as 
\begin{align}
&\bB = B_0(z) \hat z \nonumber \\
&\hspace{5 pt}+ \bigg[ \frac{B_1(z)}{R} - \frac{B_0'(z)}{2} \bigg] x \hat x + \bigg[ - \frac{B_1(z)}{R} - \frac{B_0'(z)}{2} \bigg] y \hat y , 
\end{align}
where $\hat x$, $\hat y$, and $\hat z$ are the unit vectors corresponding to the usual $(x,y,z)$ Cartesian coordinates. 

Consider a field line that passes through the point $(x_0, y_0, z_0)$. 
If the field line is parameterized by $z$, we can write that the trajectory of the field line satisfies
\begin{align}
&\frac{\D x_F}{\D z} = \frac{x_F B_1}{R B_0} - \frac{x_F B_0'}{2 B_0} \\
&\frac{\D y_F}{\D z} = - \frac{y_F B_1}{R B_0} - \frac{y_F B_0'}{2 B_0} \, ,
\end{align}
which can be solved to get 
\begin{align}
x_F(z) &= x_0 \bigg[ \frac{B_0(z)}{B_0(z_0)} \bigg]^{-1/2} \exp \bigg[ + \int_{z_0}^z \frac{B_1(z) \, \D z}{R B_0(z)} \bigg] \label{eqn:fieldX} \\
y_F(z) &= y_0 \bigg[ \frac{B_0(z)}{B_0(z_0)} \bigg]^{-1/2} \exp \bigg[ - \int_{z_0}^z \frac{B_1(z) \, \D z}{R B_0(z)} \bigg]. \label{eqn:fieldY}
\end{align}
Let 
\begin{gather}
\xi \doteq \int_{z_0}^z \frac{B_1(z) \, \D z}{R B_0(z)} \, . \label{eqn:xi}
\end{gather}
Then with $r^2 = x_F^2 + y_F^2$ and $r_0^2 \doteq x_0^2 + y_0^2$, the radial position of the field line at axial positions $z_0$ and $z$ can be related by 
\begin{align}
r_0^2 = r^2 \bigg[ \frac{B_0(z)}{B_0(z_0)} \bigg] \bigg[ \cos^2 (\theta) e^{-2 \xi} + \sin^2(\theta) e^{+2 \xi} \bigg]. 
\end{align}
This means that the electrostatic potential $\varphi_\text{iso}$ can be written explicitly as 
\begin{align}
&\varphi_\text{iso}(r, \theta, z) = \nonumber \\
&- \frac{E_0 r^2}{2 R} \bigg[ \frac{B_0(z)}{B_0(z_0)} \bigg] \bigg[ \cos^2 (\theta) e^{-2 \xi} + \sin^2(\theta) e^{+2 \xi} \bigg]. 
\end{align}
Consider the case in which the guide field $B_0$ is constant. 
Moreover, suppose $\xi \ll 1$. 
Then 
\begin{align}
\varphi_\text{iso} = - \frac{E_0 r^2}{2 R}  \bigg[ 1 - 2 \xi \cos(2 \theta) + 2 \xi^2 + \mathcal{O} \big( \xi^3 \big) \bigg]. \label{eqn:quadrupolePhiLeadingOrder}
\end{align}
This results in lab-frame fields 
\begin{align}
E_r &= \frac{E_0 r}{R} \bigg[ 1 - 2 \xi \cos(2 \theta) + 2 \xi^2 + \mathcal{O}\big( \xi^3 \big) \bigg] \label{eqn:quadrupoleEr} \\
E_\theta &= \frac{2 E_0 r}{R}  \bigg[  \xi \sin (2 \theta) + \mathcal{O} \big(\xi^3\big) \bigg] \label{eqn:quadrupoleEtheta} \\
E_z &= - \frac{E_0 r^2}{R} \bigg[ \frac{\D \xi}{\D z} \big( \cos (2 \theta) - 2 \xi \big) + \mathcal{O}\big( \xi^3 \big) \bigg]. \label{eqn:quadrupoleEz}
\end{align}
In regions where $B_1(z)$ does not change quickly, $E_z$ is comparatively small so long as $r$ is small compared with $z - z_0$ (though note that the assumption that $\xi \ll 1$ implicitly means that $z - z_0$ must itself be small compared with $R B_0 / B_1$). 

\subsection{Higher-$n$ Multipole Fields}

The exact solution for $\varphi_\text{iso}$ found when $n = 2$ is less tractable for $n>2$. 
However, it turns out that up to the same $\mathcal{O}(\xi^3)$ corrections neglected when $n = 2$, the fields can be found for any $n$. 
Consider the same fields as before, with $B_0 = \text{constant}$ and now with some integer $n \geq 2$. 
One can make the following ansatz with the same form as Eq.~(\ref{eqn:quadrupolePhiLeadingOrder}): 
\begin{gather}
\varphi_\text{iso} = - R E_0 \bigg[ \frac{r^2}{2 R^2} - \frac{\xi r^n}{R^{n}} \cos(n \theta) + \frac{n r^{2n-2} \xi^2}{2 R^{2n-2}} + \mathcal{O}(\xi^3) \bigg] . 
\end{gather}
This leads to 
\begin{align}
E_r &= \frac{E_0 r}{R} \bigg[ 1 - \frac{n r^{n-2} \xi}{R^{n-2}} \cos(n \theta) + \frac{n(n-1) r^{2n-3} \xi^2}{R^{2n-3}} \nonumber \\
&\hspace{50 pt}+ \mathcal{O}(\xi^3) \label{eqn:ErApprox} \bigg] \\
E_\theta &= \frac{n E_0 r^{n-1}}{R^{n-1}} \bigg[ \xi \sin(n \theta) + \mathcal{O}(\xi^3) \bigg] \label{eqn:EthetaApprox} \\
E_z &= - \frac{E_0 r^n}{R^{n-1}} \bigg[ \frac{\D \xi}{\D z} \bigg( \cos(n \theta) - \frac{n r^{n-2} \xi}{R^{n-2}} \bigg) + \mathcal{O}(\xi^3) \bigg] \label{eqn:EzApprox} , 
\end{align}
with $\xi$ still defined as in Eq.~(\ref{eqn:xi}). 
Note that these expressions match Eqs.~(\ref{eqn:quadrupoleEr}-\ref{eqn:quadrupoleEz}) when $n = 2$. 
These fields satisfy 
\begin{align}
\bE \cdot \bB &= n (n-1) \frac{r^{3n-2}}{R^{3n-2}} E_0 B_1 \xi^2 \cos(n \theta) + \mathcal{O}(\xi^3). 
\end{align}
If the gradient scale length of $\xi$ is long compared with $r$, then $B_1 / B_0$ is as small or smaller than $\mathcal{O}(\xi)$. 

The corresponding fields in the frame rotating at $\rot$ can be written up to $\mathcal{O}(E_0 \xi^3)$ corrections as 
\begin{align}
E_{*r} &= - \frac{n E_0 r^{n-1}}{R^{n-1}} \bigg[ \xi \cos(n \rot t + n \theta_*) - \frac{(n-1) r^{n-1} \xi^2}{R^{n-1}} \bigg] \\
E_{*\theta} &= + \frac{n E_0 r^{n-1}}{R^{n-1}} \bigg[ \xi \sin(n \rot t + n \theta_*)  \bigg] \\
E_{*z} 
&= \frac{(\omega_E - \rot) B_1 r^n}{R^{n-1}} \cos(n\theta) - \frac{n \omega_E B_1 r^{2n-2}}{R^{2n-3}} \xi . 
\end{align}
The rotating-frame fields have both parallel and perpendicular components. 
Note that $E_{*r}$ and $E_{* \theta}$ are (1) oscillatory at frequency $n \omega_\text{rot}$ and (2) identical up to a $\pi/2$ phase shift and $\mathcal{O}(\xi^2)$ corrections. 
If a particle has approximately constant $r_*$ and $\theta_*$ in the rotating frame, then perpendicular components of the time-varying fields in that frame are \textit{circularly polarized}. 
In the notation of Eq.~(\ref{eqn:EMinus}), this is the $E_-$ polarization with frequency $n \omega_\text{rot}$. 

The magnitude of the parallel-polarized component of $\bE_*$ can be found by taking the dot product with $\bB_*$. Up to higher-order corrections in $\xi$, we get 
\begin{align}
	E_{*||} &= \frac{B_1 (\omega_E - \rot) r^n}{R^{n-1}} \cos(n \rot t + n \theta_*) \nonumber \\
	&\quad- \frac{n B_1 \omega_E r^{2n-2}}{R^{2n-3}} \frac{2 \rot}{\Omega_0 + 2 \rot} \, \xi . \label{eqn:EStarParallel}
\end{align}
The sign of the first term in Eq.~(\ref{eqn:EStarParallel}) depends on the correlation of $r$ with $\theta$; see Appendix~\ref{sec:orbits}. 
The second term is negative if $E_0$, $B_1$, and $\xi$ are positive. 
The square magnitude of $\bE_*$ is 
\begin{align}
|E_*|^2 &= \bigg[ \frac{n E_0 r^{n-1}}{R^{n-1}} \xi \bigg]^2 \nonumber \\
&\hspace{10 pt}+ \bigg[ \frac{B_1 (\omega_E - \rot) r^n}{R^{n-1}} \cos^2 (n \rot t + n \theta_*) \bigg]^2, 
\end{align}
so we can see that the smallness of the perturbation enters into $E_{*||}$ through $B_1 / B_0$, whereas it enters into $E_*$ through $\xi = \int_{z_0}^z B_1 \D z / R B_0$. 
As such, $E_{*||} \ll E_*$ whenever $B_1 / B_0 \ll \xi$ -- or, equivalently, whenever the gradient scale length of $\xi$ is long compared with $R$. 
In these cases, the circularly-polarized component of $\bE_*$ is much larger than the parallel-polarized component, and the ponderomotive potential is well-approximated by the contribution from the circular-polarized component alone. 

However, this does not mean that the parallel fields can be neglected. 
Although $E_{||} = 0$ in the laboratory frame, and although a significant part of $E_{*||}$ is oscillatory, $E_{*||}$ does not average to zero over an oscillation. 
A field component that does not average to zero can contribute to the dynamics at a lower order than a comparably strong field component that does, because it can act on particles through direct acceleration rather than through the ponderomotive effect. 

So, when does this direct acceleration compete with the ponderomotive effect? 
Note that both terms in Eq.~(\ref{eqn:EStarParallel}) contribute a nonzero term, on average. 
The first contributes $\mathcal{O}(\xi)$ corrections to $r$ over the course of an oscillation; see Appendix~\ref{sec:orbits}. 
The second, of course, is simply not oscillatory. 
The contribution from the nonoscillatory part of $E_{*||}$ comes from 
\begin{gather}
\Delta W_\text{direct} \approx \int q E_{*||} \D z, 
\end{gather}
since $\bB_*$ is approximately parallel to $\hat z$. 
This amounts to the non-conservation of the form of the particle's ``quasi-energy'' given in Eq.~(\ref{eqn:quasiEnergy}). 
Keeping in mind that $B_1 / R B_0 = \D \xi / \D z$, both of the two terms in Eq.~(\ref{eqn:EStarParallel}) contribute $\mathcal{O}(\xi^2)$ terms to $\Delta W_\text{direct}$. 
This can be understood, at least in part, from the fact that the Coriolis force leads to a gap between flux surfaces and particle drift surfaces. 
As the magnetic field geometry guides a particle inwards or outwards, the Coriolis force systematically pushes the particle radially to regions with lower potential (electrostatic) energy. 
This is closely related to a mechanism of cross-field conductivity that appears in rotating systems.\cite{Rax2019, Kolmes2019} 

Meanwhile, the ponderomotive potential from the circularly-polarized part of $\bE_*$ is 
\begin{align}
\Phi_* &= m \frac{\Omega_0^2 \omega_E^2}{4 n \rot} \frac{1}{(n-2) \rot - \Omega_0} \frac{n^2 r^{2n-2}}{R^{2n-4}} \xi^2. 
\end{align}
The key takeaways here are as follows. 
First, $E_{*||}$ is nonzero but appears at higher order in $\xi$ than the circularly-polarized component of $\bE_*$. 
Therefore, the ponderomotive potential is dominantly set by the circularly-polarized wave that appears in the rotating frame rather than the parallel-polarized wave. 
Second, even though $E_{*||}$ is small, it does not average to zero. 
This means that (at leading order in $\xi$), $E_{*||}$ does not contribute to the ponderomotive potential $\Phi_*$, but it \textit{does} contribute a direct acceleration term $\Delta W_\text{direct}$ that appears at the same order as $\Phi_*$. 
Third, although $\Phi_*$ and $\Delta W_\text{direct}$ are both $\mathcal{O}(\xi^2)$, $\Delta W_\text{direct}$ is suppressed relative to $\Phi_*$ in cases where $\rot \ll \Omega_0$. 
This does not include all cases of interest, but it is the limit typically envisioned for fusion devices.

\begin{figure*}
\centering
\hspace{-10pt}
\includegraphics[scale=0.5]{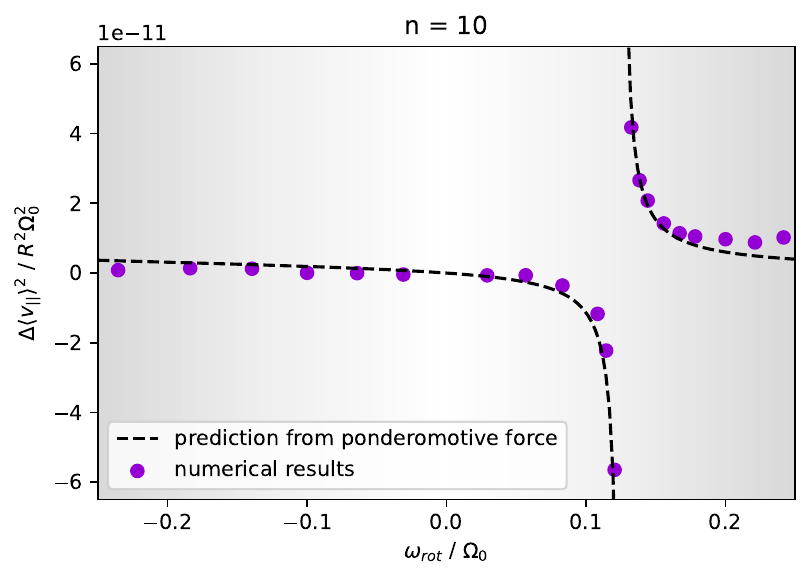} \quad
\includegraphics[scale=0.5]{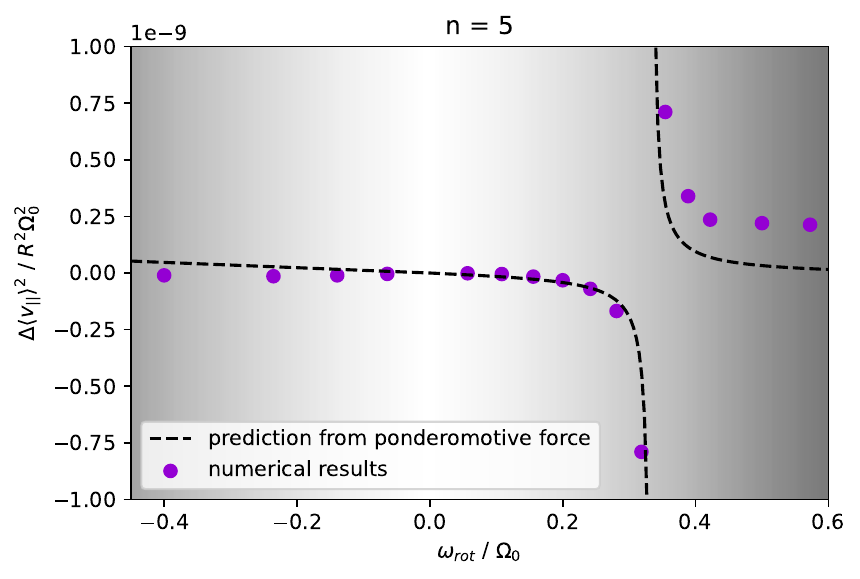}
\caption{Comparison of the effective potential barrier at $z = \ell$ between the low-frequency analytic predictions and numerics obtained using a particle-pusher. The analytics use the expression for the ponderomotive potential using the approximation that the field in the rotating frame is purely circularly polarized; see Eq.~(\ref{eqn:ponderomotivePhi}). As expected, we find that the circularly-polarized ponderomotive potential is a good approximation for the dynamics at low $\rot / \Omega_0$ but a weaker approximation at higher $\rot$. The shaded backgrounds provide a reference scale for the validity of the approximation, with darker gray appearing when $\rot$ becomes less small compared with $\Omega_0$. } 
\label{fig:numerics}
\end{figure*}

These basic conclusions are straightforward to test numerically. 
Consider, for example, the field amplitude used by \citet{Ochs2023Isopotentials}: 
\begin{gather}
B_1(z) = B_{10} \frac{a (z / \ell)^{a-1}}{[1+(z/\ell)^a]^2} \, , 
\end{gather}
where $\ell$ is a characteristic length-scale parameter and $a$ is a constant.  
This can be integrated to get 
\begin{gather}
\xi = \frac{1}{R B_0} \int_0^z B_1(z') \, \D z' = \frac{B_{10}}{B_0} \frac{(\ell / R) (z/\ell)^a}{1+(z/\ell)^a} \, . 
\end{gather}
Figure~\ref{fig:numerics} shows results using these magnetic fields with $\ell / R = 5$, $B_{10} / B_0 = 10^{-5}$, and $a = 5$. 
The electric fields are taken from Eqs.~(\ref{eqn:ErApprox}-\ref{eqn:EzApprox}). 
Particles are initiated at $(x,y,z) = (4 R / 5,0,0)$ and $\bv = .003 R \Omega_0 \hat z + (4 R / 5) \rot \hat \theta$. 
This is equivalent to purely parallel propagation in the rotating frame, so effects like mirroring should not appear. 
The simulations were performed with a range of values of $\rot$ (set by varying the electric field). 
For each simulation, the change in $(m/2) \langle v_{||} \rangle^2$ was calculated at $z = \ell$, with the averaging performed using a 1D uniform filter implemented in the \texttt{scipy} package.\cite{scipy} 
Orbit integration was performed using the Zenitani-Umeda\cite{Zenitani2018} second-order Boris pusher\cite{Qin2013} generalization, implemented in the \texttt{ploopp} code. 

The simulations confirm our expectations: when $|\rot / \Omega_0|$ is less than about 0.2, the results closely match the behavior from Eq.~(\ref{eqn:ponderomotivePhi}). 
For larger $|\rot / \Omega_0|$, the numerics noticeably deviate from the behavior of the circularly-polarized ponderomotive potential. 

In general, $E_{*||}$ may take either sign; depending on the parameters of the system, the parallel acceleration can either increase or reduce the effects of the ponderomotive potential. 
In the examples shown in Figure~\ref{fig:numerics}, its most noticeable effect seems to be to reinforce the ponderomotive potential when $(n-2) \rot > \Omega_0$ (to the right of the resonance).

\section{Discussion} \label{sec:discussion}

In a magnetostatic ponderomotive configuration, a ponderomotive effect is produced by causing the plasma to flow across a static field perturbation (rather than introducing a time-dependent wave).\cite{Rubin2023} 
\citet{Ochs2023Isopotentials} showed in a slab model that it is sometimes possible to calculate the effective potential very easily by writing down the fields in the co-moving frame, calculating the polarization of the wave that appears in that frame, and comparing with Eq.~(\ref{eqn:ponderomotivePhi}). 
Here we have shown how this picture needs to be modified when considering a rotating cylindrical system rather than a slab. 

Some of these modifications are relatively straightforward. 
The rotating-frame transformation affects fields differently from the translating-frame transformation, and this modifies the form of the ponderomotive potential. 
This is shown in Eq.~(\ref{eqn:ponderomotivePhiStar}). 
Implications include the disappearance of one of the ponderomotive resonances when $n = 2$ (that is, for a quadrupole perturbation). 

Other modifications are more subtle. 
Because the rotating-frame transformation is non-inertial, $\bE \cdot \bB$ need not be preserved. 
This means that a system with vanishing $E_{||}$ in the laboratory frame may have a nonzero parallel electric field in the rotating frame. 
As we show in the example of a multipole configuration, the effects of this field can be important even when the rotating-frame $E_{*||}$ is a small part of the total electric field. 
This is because even an oscillatory perturbation can give rise to a parallel electric field in the rotating frame that does not average to zero. 

In the course of identifying these inertial effects, we also identify the conditions under which they can be suppressed. 
The key parameter that must be small in order to neglect inertial effects is the ratio of the rotation frequency to the gyrofrequency. 
When this parameter is small, the slab-like limit is recovered and the behavior of the ponderomotive dynamics once again becomes relatively simple. 
Of course, all of the above considerations could help to inform the design of devices that rely on these effects (for example, how to tune perturbations in order to achieve a desired ponderomotive potential, and in what parameter regime a potential could be effective). 

The example presented in Section~\ref{sec:multipoleExample} can also help to clarify the relationship between the cylindrical system considered by \citet{Rubin2023, Rubin2023ii} and the slab system considered by \citet{Ochs2023Isopotentials}. 
Our results confirm the general point that changing the isopotential surfaces from the unperturbed to the perturbed magnetic field lines does entirely change the polarization; recall that the unperturbed-field-line system used in \citet{Rubin2023, Rubin2023ii} had a dominantly parallel-polarized ponderomotive barrier. 
However, we find here that the perturbed-isopotential version of the \citet{Rubin2023, Rubin2023ii} system results in approximately circular polarization, whereas the perturbed-isopotential version of the \citet{Ochs2023Isopotentials} system was linearly polarized. 
This makes sense; perturbed-field-line isopotential surfaces should always lead to some combination of the $E_+$ and $E_-$ polarizations (up to the inertial corrections discussed above), but the mix of $E_+$ and $E_-$ will depend on the particular choice of fields. 
The appearance of the linear transverse polarization in \citet{Ochs2023Isopotentials} and circular polarization in Section~\ref{sec:multipoleExample} reflects a difference in the structure of these fields. 

Our focus here has been exclusively on the question of how to understand the relationship between the fields and the resulting ponderomotive force (and specifically the role that inertial forces play in that relationship). 
There is, of course, another important problem that is closely related: how to determine what fields can actually be imposed on a real plasma system using practical coils or electrodes. 
This is a separate problem, and is discussed in more detail elsewhere.\cite{Rubin2024} 
In fact, it appears that the most promising practical candidate may be an X mode driven by electrostatic perturbations (as opposed to the magnetostatic perturbations envisioned in the example in Section~\ref{sec:multipoleExample}).\cite{Rubin2024} 
These issues are related more generally to the problem of wave propagation in rotating plasmas. \cite{Lehnert1954, Lehnert1955, Tandon1966, Uberoi1970, Verheest1974, Rax2021, Gueroult2023, Rax2023QL, Rax2023AW, Langlois2023}  
That being said, there is one comment on the wave-propagation problem that is worth making here. 
If we are in a limit where we expect $E_{||}$ to vanish -- for example, due to the high mobility of charge carriers along field lines -- it is reasonable to wonder whether it is the lab-frame $E_{||}$ or the rotating-frame $E_{*||}$ that ought to vanish. 
Note that the rotating-frame transformation depends on the charge and mass of the particle species being considered. 
The electrons are typically the most mobile charge carrier, and their low mass means that there is little difference between the parallel field in the different electron frames for them. 
In other words, if we are considering a limit where high charge-carrier mobility prevents parallel fields, it is sensible to take the laboratory-frame $E_{||}$ to vanish and to allow the \textit{ion} rotating-frame $E_{*||}$ to be nonzero. 

Another question that we have not resolved here relates to the dynamics in the regime where the ponderomotive effect is not dominant. 
We have identified the limit within which the ponderomotive potential $\Phi_*$ sets the dynamics, and we have shown how rotation modifies $\Phi_*$, but we have not calculated the dynamics in the regime in which $\Phi_*$ does not dominate. 
This goes beyond our intended scope here, but would be an interesting issue to explore in the future. 

All of the physics considered here is nonresonant, and indeed much of the analysis would not be valid too close to a resonance.\cite{Guy1982} 
Static field perturbations can also be used to produce resonant wave-particle interactions; this was proposed by \citet{Fetterman2010} in the context of $\alpha$ channeling.\cite{Fisch1992} 
The structure of the perpendicular-polarized ponderomotive force leads to stronger forces close to resonances, that is, as $-(n+2) \rot$ or $(n-2) \rot$ approach $\Omega_0$. 
Correctly simulating the behavior very near resonance can be a subtle problem, and is likely to benefit from advanced computational techniques.\cite{Qin2024} 
For applications in which resonant effects are undesirable but strong ponderomotive forces are desirable, this leads to an interesting situation in which it is optimal to get as close to a resonance as possible without actually producing resonant interactions. 

Also note that a helical magnetic perturbation could produce an effective wave with a nonzero wavenumber component in the parallel direction. 
One could consider, for example, the helical structure imposed on the SMOLA device at the Budker Institute.\cite{Inzhevatkina2021} 
Such a perturbation would duly modify the potential described in Eq.~(\ref{eqn:ponderomotivePhiStar}); the resulting Doppler shift would depend on $v_{||}$, the parallel component of the particle's velocity. 

Of course, when considering a species-dependent effect like this one, it is important to keep in mind that the overall response of a plasma system will include both the effect itself and the self-consistent response of the plasma to the effect. 
A helpful analogy might be the trapping effect in a magnetic mirror: magnetic mirrors more effectively trap ions than electrons, so the imposition of a mirror field leads to an ambipolar electric field that tends to equalize the overall confinement times for ions and electrons. 
When considering the behavior of the plasma as a whole, it is sometimes practical to work with a combined ponderomotive force, summed over multiple species.\cite{Browning1988, Browning1989} 
This would be reasonably straightforward to do for the case considered here, though it would be important to keep in mind that parameters like the rotation frequency are also species-dependent.


\begin{acknowledgements}

This work has benefited from helpful conversations with Alex Glasser, Mike Mlodik, Ian Ochs, and Tal Rubin. 

This work was supported by ARPA-E Grant No. DE-AR0001554. 
This work was also supported by the DOE Fusion Energy Sciences Postdoctoral Research Program, administered by the Oak Ridge Institute for Science and Education (ORISE) and managed by Oak Ridge Associated Universities (ORAU) under DOE Contract No. DE-SC0014664. 
	
\end{acknowledgements}

\section*{Data Availability Statement}

Data sharing is not applicable to this article as no new data were created or analyzed in this study.

\section*{Author Declarations}

The authors have no conflicts to disclose. 

\bibliographystyle{apsrev4-2} 
\bibliography{../../../Dropbox/Master.bib}

\appendix 

%
%
%
%
%

\newpage
\section{Approximate Orbit Calculation} \label{sec:orbits}

In Section~\ref{sec:multipoleExample}, in the discussion of the rotating-frame parallel field $E_{*||}$, it is helpful to have a sense for what the particle orbits behave, and particularly how $r(t)$ might vary as a function of time. 

The simplest scenario to consider is one in which $\xi$ is constant and $B_1 / B_0 \ll \xi$. 
Either of these assumptions could be worth revisiting in a more detailed calculation of the dynamics. 

The equations of motion are 
\begin{align}
	\ddot r - r \dot \theta^2 &= \frac{q}{m} \big( E_r + v_\theta B_z - v_z B_\theta \big) \\
	r \ddot \theta + 2 \dot r \dot \theta &= \frac{q}{m} \big( E_\theta - v_r B_z + v_z B_r \big) \\
	\ddot z &= \frac{q}{m} \big( E_z + v_r B_\theta - v_\theta B_r \big). 
\end{align}
Consider an expansion of $r$ and $\theta$ into $r_i$ as follows: 
\begin{gather}
r = \sum_{i=0}^\infty r_i \\
\theta = \sum_{i=0}^\infty \theta_i, 
\end{gather}
where $r_i, \theta_i \sim \mathcal{O}(\xi^i)$. 
The leading-order solution has $r_0 = \text{constant}$ and $\theta_0 = \rot t$. 

Then, setting $\rho \doteq r_0 / R$, the $\mathcal{O}(\xi)$ equations of motion are 
\begin{align}
\ddot r_1 &= \Omega_0 \bigg[ n \omega_E \rho^{n-1} \xi \cos(n \rot t) + r_0 \dot \theta_1 \bigg( 1 + \frac{2 \rot}{\Omega_0} \bigg) \bigg] \\
r_0 \ddot \theta_1 &= \Omega_0 \bigg[ - n \omega_E \rho^{n-1} \xi \sin(n \rot t) - \dot r_1 \bigg( 1 + \frac{2 \rot}{\Omega_0} \bigg) \bigg] . 
\end{align}
One solution (with initial conditions chosen to suppress the gyromotion) has 
\begin{gather}
r_1 = \frac{\Omega_0 \omega_E}{\rot [ \Omega_0 - (n-2) \rot ]} \frac{r^{n-1}}{R^{n-2}} \xi \cos(n \rot t). 
\end{gather}
For present purposes, the important thing to point out is that to first order in $\xi$, 
\begin{align}
r^n \cos(n \rot t) = (r_0^2 + n r_0 r_1) \cos(n \rot t) + \mathcal{O}(\xi^2), 
\end{align}
which is nonzero at $\mathcal{O}(\xi)$ when averaged over a $\rot^{-1}$ timescale. 
This is significant for the interpretation of Eq.~(\ref{eqn:EStarParallel}). 

\end{document}